\documentclass[conference]{IEEEtran}
\IEEEoverridecommandlockouts
\usepackage{cite}
\usepackage{amsmath,amssymb,amsfonts}
\usepackage{algorithmic}
\usepackage{subfig}
\usepackage{graphicx}
\usepackage{textcomp}
\usepackage{xcolor}
\usepackage{pifont}
\usepackage{enumerate}
\usepackage{booktabs}
\usepackage{multirow}
\def\BibTeX{{\rm B\kern-.05em{\sc i\kern-.025em b}\kern-.08em
    T\kern-.1667em\lower.7ex\hbox{E}\kern-.125emX}}

\newtheorem{mydef}{Definition}
    
\begin{document}

\title{Sign-Guided Bipartite Graph Hashing for Hamming Space Search\\
}

\author{\IEEEauthorblockN{Xueyi Wu}
\IEEEauthorblockA{\textit{East China Normal University} \\
xueyi.wu.1@stu.ecnu.edu.cn}
}

\maketitle

\begin{abstract}

Bipartite graph hashing (BGH) is extensively used for Top-$K$ search in Hamming space at low storage and inference costs. 
Recent research adopts graph convolutional hashing for BGH and has achieved the state-of-the-art performance. However, the contributions of its various influencing factors to hashing performance have not been explored in-depth, including the same/different sign count between two binary embeddings during Hamming space search (sign property), the contribution of sub-embeddings at each layer (model property), the contribution of different node types in the bipartite graph (node property), and the combination of augmentation methods. 
In this work, we build a lightweight graph convolutional hashing model named LightGCH by mainly removing the augmentation methods of the state-of-the-art model BGCH.
By analyzing the contributions of each layer and node type to performance, as well as analyzing the Hamming similarity statistics at each layer, we find that the actual neighbors in the bipartite graph tend to have low Hamming similarity at the shallow layer, and all nodes tend to have high Hamming similarity at the deep layers in LightGCH. To tackle these problems, we propose a novel sign-guided framework SGBGH to make improvement, which uses sign-guided negative sampling to improve the Hamming similarity of neighbors, and uses sign-aware contrastive learning to help nodes learn more uniform representations. Experimental results show that SGBGH outperforms BGCH and LightGCH significantly in embedding quality.

\end{abstract}


\section{Introduction}

Bipartite graphs are widely used for modeling the interactions between two different types of entities in real-world applications, such as search engines~\cite{web_search_vldb,web_search_sigmod,pvldb/AntonellisGC08} and recommender systems~\cite{MultiBiSage,HBGNN,Pixie,PinSage,commodity.embedding}. Top-$K$ search is a fundamental task to support these applications. For example, in the search engine, when given a query keyword, Top-$K$ search should return its most related $K$ web pages. In recent works, learning vectorized representations (i.e., embeddings) for graph nodes to boost the search task has aroused many research interests, because the candidate list can be obtained through simple operations like ranking the dot products of node embeddings.

However, with the growth of graph node count, the commonly used floating-point node embeddings can lead to substantial storage requirements, and online inference incurs lots of floating-point operations. In order to solve these problems, many works apply \textit{learning to hash}~\cite{survey.learning.to.hash} to reduce both storage and inference overheads, therefore Top-$K$ search can be conducted in the Hamming space, which is known as bipartite graph hashing (BGH).

Early BGH solutions aim to learn binary embeddings for each node based on matrix decomposition~\cite{BCCF,PPH,DFM} or discrete optimization~\cite{DCF,DPR,MFDCF}. While with the development of machine learning technology and the significant success of graph neural networks (GCN) in bipartite graph embedding~\cite{NGCF,LightGCN,BiGI,SHT,HCCF}, recent BGH researches tend to focus on \textit{expressive embedding form}, \textit{powerful model} and \textit{training convenience}. For example, they tend to introduce the \textbf{mixed-precision embedding} (a combination of floating-point rescaling factor and binary embedding in Fig.~\ref{fig:bin_emb}) to enhance embedding expressiveness~\cite{BiGeaR,BGCH}, adopt various GCNs to capture bipartite graph structure~\cite{HashGNN,HS-GCN,BiGeaR,BGCH,BinCF,GCNH}, and employ end-to-end training methods to simplify training process~\cite{HashGNN,HS-GCN,CIGAR,BGCH}. Among them, BGCH~\cite{BGCH} is the state-of-the-art end-to-end solution due to its joint improvements in the above three aspects.

Although BGCH provides significant reference for the design of graph convolutional hashing model, the contributions of its various influencing factors on performance have not been explored in-depth. Therefore, there is still room for improvement in BGH. For instance, (1) For \textit{embedding form}, there are two decisive factors for dot product computation in Top-$K$ search for the mixed-precision embedding: the rescaling factor and the same/different sign count between two binary embeddings (\textit{sign property}). The latter is critical for Hamming space search but is rarely discussed in related work. (2) For \textit{model property}, although GCN is widely used in bipartite graph embedding, \cite{LayerGCN} points out different layers of GCN exhibit large differences. Hence, it is necessary to explore the contribution of sub-embeddings at each layer of graph convolutional hashing model like BGCH, since it concatenates sub-embeddings of each layer to generate  final embeddings. (3) For \textit{node property}, two node types contribute differently to downstream tasks, and GCN will further make different types of nodes exhibiting varying performance across different layers, which calls for further analysis and elaborate processing. (4) For \textit{augmentation methods} like feature dispersion and other training techniques in BGCH, although their effectiveness is validated in individual ablation study, their combined effect on the overall improvement has not been studied.

Next, we dig into the above four factors, to discuss the challenges in BGH:

\begin{figure}
    \centering
    \includegraphics[width=\linewidth]{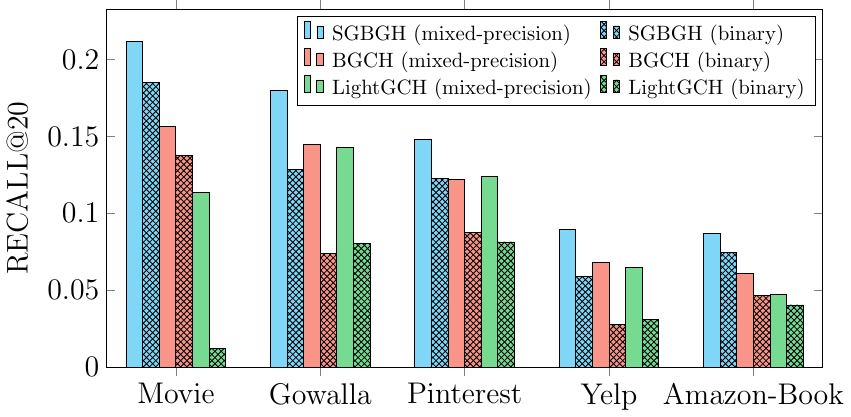}
    \caption{Performance comparison of mixed-precision embedding and binary embedding of different hashing methods. SGBGH and LightGCH are both our proposed methods.}
    \label{fig:BGCH_bin}
\end{figure}

\noindent\textbf{Challenge \uppercase\expandafter{\romannumeral1}:} \textit{How to make better use of the sign property? } Since the dot product computation is indispensable in Top-$K$ search, BGH performance is influenced by the rescaling factor and the same/different sign count between two binary embeddings under the mixed-precision embedding form. When the rescaling factor remains unchanged, the more same sign count two embeddings have, the larger their dot product value. We term this as the \textbf{sign property} of learning to hash. The sign property is important since it ensures the most basic performance, as we observe that the mere binary embeddings can already guarantee about 67\% performance of the original mixed-precision embeddings in BGCH as shown in Fig.~\ref{fig:BGCH_bin}. However, there is little exploration for the sign property except for~\cite{HS-GCN}. This observation inspires us to improve BGH based on the sign property.

\noindent\textbf{Challenge \uppercase\expandafter{\romannumeral2}:} \textit{How to propose a more general augmentation method?} We first build a \underline{Light}weight \underline{G}raph \underline{C}onvolutional \underline{H}ashing model called LightGCH by mainly removing augmentation methods such as feature dispersion and reconstruction loss from BGCH. By comparing LightGCH and BGCH in Fig.~\ref{fig:BGCH_bin}, we observe that their performance is quite close in terms of both mixed-precision embeddings and binary embeddings on three datasets (Gowalla, Pinterest and Yelp). This suggests that the combination of different augmentation methods may offset each other's effects, leading to a relatively limited overall improvement. Therefore, proposing a more general augmentation method is necessary.

\begin{figure}
	\centering
	\begin{minipage}[t]{\linewidth}
	\centering
        \subfloat[performance contribution of each layer]{
		\centering
		\includegraphics[width=0.46\linewidth]{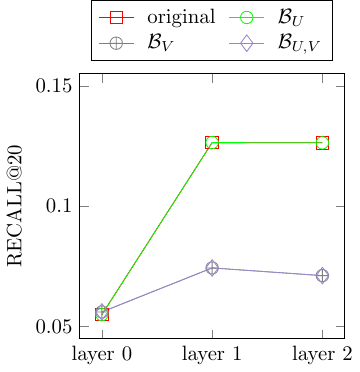}
		\label{fig:103.LightGCH_bin_group_layer}
	}
        \hspace{1mm}
        \subfloat[performance contribution of each node type]{
		\centering
            \includegraphics[width=0.45\linewidth]{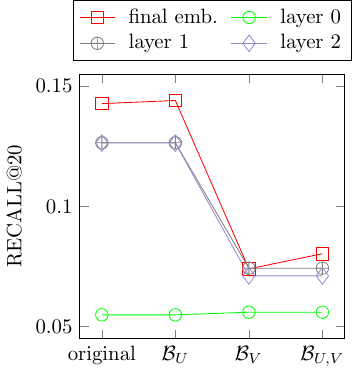}
		\label{fig:102.LightGCH_bin_group_UV}
	}
	\centering
	\caption{Performance contribution of each node type at each layer in LightGCH on Gowalla dataset. $\mathcal{B}_U$, $\mathcal{B}_V$, $\mathcal{B}_{U,V}$ denote binarizing embeddings of nodes in $U$, $V$, $U \cup V$ per layer.} \label{fig:LightGCH_bin}
\end{minipage}%
\end{figure}

\noindent\textbf{Challenge \uppercase\expandafter{\romannumeral3}:} \textit{How to analyze a graph convolutional hashing model based on model and node properties, identify the latent issues, and make improvement?} Although BGCH's augmentation methods like feature dispersion and reconstruction loss  act at the shallow layer, their influence diffuse across all layers due to the graph convolution. Therefore,  the study of model property should also consider the indirect effects of augmentation methods. Thanks to LightGCH in \textbf{Challenge \uppercase\expandafter{\romannumeral2}}, we can take it as a simplified example for further analysis since it shares similar performance with BGCH in most cases. We select Gowalla dataset and conduct binarization experiments on a 2-layer LightGCH using binarization function $\operatorname{sign}(\cdot)$. Observe from Fig.~\ref{fig:LightGCH_bin} that the performance trends of shallow embeddings (at layer 0) and deep embeddings (at layers 1 and 2) under various binarization scenarios are almost inconsistent. This indirectly indicates different properties of the shallow layer and the deep layers. Understanding the reasons behind this phenomenon and uncovering potential issues are crucial for model improvement.

To address the above challenges, we investigate LightGCH at different layers from a novel perspective based on the sign property, identify distinct issues in its shallow and deep layers, and propose a novel \underline{S}ign-\underline{G}uided \underline{B}ipartite \underline{G}raph \underline{H}ashing framework SGBGH, to enhance its  performance (\textbf{Challenge \uppercase\expandafter{\romannumeral1}}). Specifically, by analyzing the contribution of each layer and each node type to performance and combining it with statistical analysis of the Hamming similarity~\cite{HS-GCN} of neighboring and non-neighboring nodes at each layer, we identify two latent issues of LightGCH: low Hamming similarity of actual neighbors at the shallow layer, and high Hamming similarity of all nodes at the deep layers (\textbf{Challenge \uppercase\expandafter{\romannumeral3}}). Then, we propose a sign-guided hard negative sampling method at the shallow layer to improve the Hamming similarity of neighboring nodes, and propose a sign-aware contrastive learning method at the deep layers to help model learn more uniform hash embeddings. By integrating these two sign-guided techniques, SGBGH alleviates the Hamming similarity issues at each layer of LightGCH, thus improving the overall effectiveness under both mixed-precision embeddings and binary embeddings, as shown in Fig.~\ref{fig:BGCH_bin} (\textbf{Challenge \uppercase\expandafter{\romannumeral2}}). To sum up, our main contributions are as follows:

\begin{itemize}
    \item We empirically show two latent issues in graph convolutional hashing model for BGH: actual neighbors have low Hamming similarity at the shallow layer, and all nodes have high Hamming similarity at the deep layers.
    \item We propose a novel sign-guided framework SGBGH to tackle the above issues, which uses sign-guided negative sampling at the shallow layer to improve the Hamming similarity of neighbors, and uses sign-aware contrastive learning at the deep layers to generate more uniform node representations.
    \item Experimental results on five real-world datasets demonstrate that SGBGH outperforms end-to-end BGH methods by a large margin in terms of hashing performance and training efficiency.
\end{itemize}

\section{Preliminaries and Problem Definition}

\subsection{Preliminaries}

\noindent\textbf{Bipartite Graph.} Given a bipartite graph $G=(U,V,E)$, $U$ and $V$ denote its two disjoint node sets, and $E\subseteq U\times V$ denotes its edge set. The nodes in $U$ and $V$ are called source nodes and destination nodes, respectively. The set of neighbors of a node $x \in U \cup V$ is denoted by $\mathcal{N}(x)$. 

\noindent\textbf{Notations.} We denote matrices in bold uppercase (e.g. $\bf{Q}$) and denote vectors in bold lowercase (e.g. $\bf{x}$). The $i$-th element of vector $\bf{x}$ is denoted by $\bf{x}\mathnormal{[i]}$.

\begin{figure}
    \centering
    \includegraphics[width=0.7\linewidth]{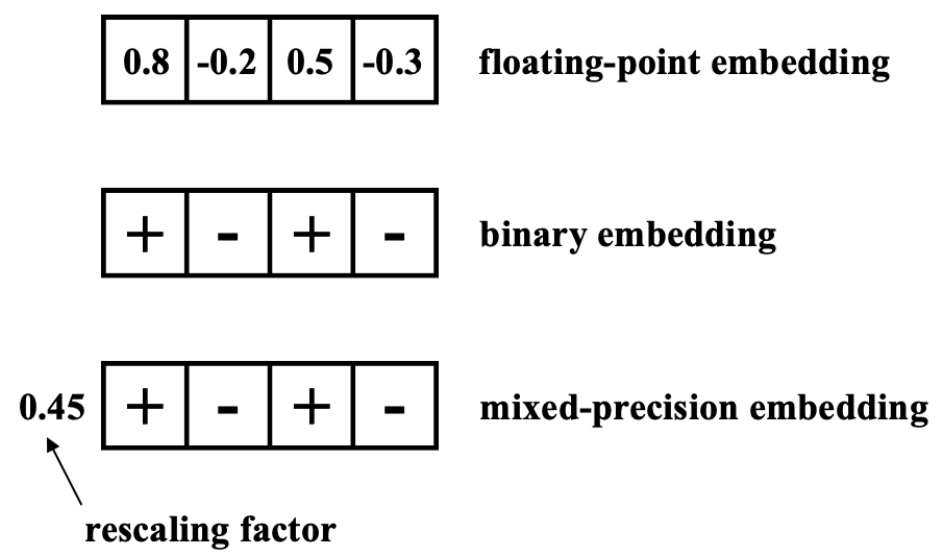}
    \caption{Floating-point embedding, binary embedding and mixed-precision embedding.}
    \label{fig:bin_emb}
\end{figure}

\noindent\textbf{Mixed-precision embedding.} Fig.~\ref{fig:bin_emb} shows a mixed-precision embedding segment. Its vectorized form is $\mathbf{q} = \alpha \mathbf{b}$, where $\alpha \in \mathbb{R}$ is the floating-point rescaling factor, and $\mathbf{b}$ is the binary embedding with each element in $\{-1,1\}$.

\noindent\textbf{Dot product computation of mixed-precision embeddings.} Given two $d$-dimensional mixed-precision embeddings $\mathbf{q}_{u} = \alpha_{u} \mathbf{b}_{u}$ and $\mathbf{q}_{v} = \alpha_{v} \mathbf{b}_{v}$, their dot product computation is as follows:
\begin{equation} \label{eq:bit_op}
\begin{aligned}
    & \left(\alpha_{u} \mathbf{b}_{u}\right)^{T} \cdot\left(\alpha_{v} \mathbf{b}_{v}\right) \\ 
    = &\alpha_{u} \alpha_{v} \mathbf{b}_{u}^{T} \mathbf{b}_{v} \\
         = &\alpha_{u} \alpha_{v} \sum\limits_\mathnormal{i\rm{=1}}^\mathnormal{d} ( \mathbf{b}_{u}[i] * \mathbf{b}_{v}[i]) \\
         = &\alpha_{u} \alpha_{v} \sum\limits_\mathnormal{i\rm{=1}}^\mathnormal{d} \left( 2 * \mathbb{I}\left( \mathbf{b}_{u}[i] = \mathbf{b}_{v}[i] \right) - 1 \right) \\
         = &\alpha_{u} \alpha_{v} \left( 2 * \operatorname{same\_sign\_count} \left( \mathbf{b}_{u}, \mathbf{b}_{v} \right) - d \right) ,
\end{aligned}
\end{equation}
where $\operatorname{same\_sign\_count}(\cdot, \cdot)$ is the same sign count of two binary embeddings, which is defined as:
\begin{equation}
\operatorname{same\_sign\_count}(\mathbf{b}_{u}, \mathbf{b}_{v}) = \sum\limits_\mathnormal{i\rm{=1}}^\mathnormal{d} \mathbb{I}\left( \mathbf{b}_{u}[i] = \mathbf{b}_{v}[i] \right). 
\end{equation}

\subsection{Problem Definition}

We aim to learn a hash function $f: U \cup V \rightarrow \mathbf{Q}$ for BGH, where the matrix $\mathbf{Q}$ denote the hash embeddings of all nodes in $U \cup V$. 
For each node $x \in U \cup V$, its hash embedding is denoted as $\mathbf{q}_{x}$. 
We aim to develop a hashing model $f(\cdot)$, so that the learned hash embeddings can be used for effective bipartite graph modeling and efficient inference in the downstream search task.

\section{Investigation of Graph Convolutional Hashing Model}
Before making improvements to the hashing model, it is essential to understand its specific structure. In this section, we will first introduce the model structure and main optimization objectives of LightGCH. By combining the contributions of each layer and node type to performance, we further analyze its latent issues from the perspective of Hamming similarity.

\subsection{LightGCH}

Fig.~\ref{fig:100.SGBGH_framework}a demonstrates the primary model structure of LightGCH, which retains only the basic lightweight graph convolutional hashing module, the layer-wise adaptive rescaling factor, the concatenated embedding form, gradient estimation method, and main optimization objectives of BGCH. Next, we will elaborate on its graph convolutional hashing module, embedding form, gradient estimation, and loss function, respectively.

\subsubsection{Lightweight Graph Convolutional Hashing Module}

This module contains two components: (1) \textit{adaptive hashing}, and (2) \textit{graph convolution}. When given the hidden representation $\mathbf{x}^{(l)} \in \mathbb{R}^{d}$ of node $x \in U \cup V$ at layer $l$, \textit{adaptive hashing} first map it to the mixed-precision form $\mathbf{q}_{x}^{(l)}$ as follows:
\begin{equation}
\begin{aligned}
    \mathbf{b}_{x}^{(l)} &= \operatorname{sign}(\mathbf{x}^{(l)}) ,  \\
    \alpha_{x}^{(l)} &= \frac{1}{d} \left\|\mathbf{x}^{(l)}\right\|_{1} ,  \\
    \mathbf{q}_{x}^{(l)} &= \alpha_{x}^{(l)} \mathbf{b}_{x}^{(l)} .
\end{aligned}
\end{equation}
Here, the mixed-precision sub-embedding of node $x$ is $\mathbf{q}_{x}^{(l)}$ at layer $l$, which is composed of the binary embedding $\mathbf{b}_{x}^{(l)}$ and the rescaling factor $\alpha_{x}^{(l)}$ obtained from L1 norm of $\mathbf{x}^{(l)}$.

\textit{Graph convolution} in LightGCH is defined as:
\begin{equation} \label{eq:hash_lgc}
\mathbf{x}^{(l+1)} = \sum\limits_{y\in \mathcal{N}(x)} \frac{1}{\sqrt{|\mathcal{N}(x)|} \sqrt{|\mathcal{N}(y)|}} \mathbf{q}_{y}^{(l)} ,
\end{equation}
which is consistent with~\cite{LightGCN} in terms of the propagation rule, while graph convolution here only takes mixed-precision embeddings as input.

\subsubsection{Final Embedding Form}

Through layer-wise adaptive hashing and graph convolution operations, node $x$ can obtain mix-precision sub-embeddings at each layer in LightGCH: $\mathbf{q}_{x}^{(0)}, \mathbf{q}_{x}^{(1)}, \cdots, \mathbf{q}_{x}^{(L)}$, where $L$ is the number of network layer. By concatenating these sub-embeddings, we have the final embedding for node $x$:
\begin{equation}
    % \mathbf{q}_{x} = \mathbf{q}_{x}^{(0)}\left\|\mathbf{q}_{x}^{(1)}\right\| \cdots \big{\|} \mathbf{q}_{x}^{(L)} .
    \mathbf{q}_{x} = \mathbf{q}_{x}^{(0)} \big{\|} \mathbf{q}_{x}^{(1)} \big{\|} \cdots \big{\|} \mathbf{q}_{x}^{(L)} .
\end{equation}

\subsubsection{Gradient Estimation}

During forward propagation of LightGCH, we strictly use $\operatorname{sign}(\cdot)$ function. During backward propagation, we estimate the gradient $\frac{\partial \operatorname{sign}(\phi)}{\partial \phi}$ through Fourier Serialized Gradient Estimation~\cite{frequency.domain.approximation} following BGCH, which is defined as:
\begin{equation}  \label{eq:fourier}
\frac{\partial \operatorname{sign}(\phi)}{\partial \phi} \doteq \frac{4}{H} \sum_{i=1,3,5, \cdots}^{n} \cos \left(\frac{\pi i \phi}{H}\right) .
\end{equation}

\subsubsection{Loss Function}

The loss function of LightGCH is:
\begin{equation}  \label{eq:hash_lightgch_objective}
\mathcal{L} = \mathcal{L}_{bpr}^{main} + \lambda \| \Theta \|^2,
\end{equation}
where $\mathcal{L}_{bpr}^{main}$ is the commonly used Bayesian Personalized Ranking (BPR) loss~\cite{BPR} in bipartite graph embedding, $\lambda$ is the regularization coefficient for model parameters. $\mathcal{L}_{bpr}^{main}$ constrains the layer-wise concatenated node embeddings:
\begin{equation}  \label{eq:hash_loss_main}
\mathcal{L}_{bpr}^{main}=\sum_{\substack{(u, v) \in E \\ v' \in N(u)}}-\log \sigma\left(\mathbf{q}_{u}^{T} \mathbf{q}_{v} - \mathbf{q}_{u}^{T} \mathbf{q}_{v^{\prime}} \right) ,
\end{equation}
where the negative sample $v'$ is obtained by the sampling method $N(\cdot)$, which is implemented as the random sampling from uniform distribution~\cite{BPR,NCF}.

\subsection{Experimental Research of LightGCH}

Since we have already demonstrated the performance trends under different binarization scenarios for each layer and node type in LightGCH in Fig.~\ref{fig:LightGCH_bin}, here, we present the preliminary observations considering the model structure.

\begin{itemize}
    \item From the perspective of network layers (Fig.~\ref{fig:103.LightGCH_bin_group_layer}), since LightGCH generates final embeddings by concatenating the mixed-precision sub-embeddings of each layer following~\cite{BGCH,BiGeaR}, rather than by averaging the sub-embeddings like~\cite{LightGCN,SGL,SimGCL,LayerGCN}, it leads to such a phenomenon: the shallow embedding (at layer 0) is concatenated into the final embedding without undergoing graph convolution operations, therefore its performance is inconsistent with deep embeddings (at layers 1 and 2) and is far inferior in embedding quality.
    \item From the perspective of node types (Fig.~\ref{fig:102.LightGCH_bin_group_UV}), the contributions of different types of nodes vary across layers of the network. In particular, the performance of shallow embeddings at layer 0 is almost identical for all binarization scenarios ($\mathcal{B}_U$, $\mathcal{B}_V$, $\mathcal{B}_{U,V}$) whether there are rescaling factors or not. On the contrary, deep embeddings at layers 1 and 2 show significant performance degradation under $\mathcal{B}_V$.
\end{itemize}

\subsubsection{Hamming Similarity}

Due to the dependence of BGH performance on the ranking of dot products, and dot products being influenced solely by two factors: the rescaling factor and the same/different sign count between binary embeddings of two types of nodes, we can analyze BGH performance by these two factors. Furthermore, we have observed that a significant issue lies in the poor performance of shallow embeddings at layer 0, which is insensitive to the value of the rescaling factor. Therefore, analyzing from the perspective of the same sign count is a feasible approach. However, it has different meanings for different embedding dimensionalities $d$ even if the same sign count remains unchanged. Hence, utilizing a universal measurement is a prerequisite for research. Here, we employ the Hamming similarity defined in HS-GCN~\cite{HS-GCN} to substitute for the same sign count for further investigation. As Hamming similarity is proportional to the same sign count and its values are within the $[0,1]$ range, researching this similarity measure can help us obtain qualitative conclusions. The definition of Hamming similarity is provided below:

\begin{mydef}[Hamming Similarity]
Hamming similarity is a similarity measure of two binary embeddings in Hamming space. Given two $d$-dimensional binary embeddings $\mathbf{b}_{u}, \mathbf{b}_{v} \in \{-1,1\}^{d}$ of nodes $u$ and $v$, their Hamming similarity is calculated as:
\begin{equation}
\begin{aligned}
\operatorname{sim}_H(\mathbf{b}_{u}, \mathbf{b}_{v}) &= \frac{1}{d} * \operatorname{same\_sign\_count} \left( \mathbf{b}_{u}, \mathbf{b}_{v} \right) \\
&= \frac{1}{d} \sum\limits_\mathnormal{i\rm{=1}}^\mathnormal{d} \mathbb{I}\left( \mathbf{b}_{u}[i] = \mathbf{b}_{v}[i] \right) . 
\end{aligned}
\end{equation}
\end{mydef}

\subsubsection{Analysis of Hamming Similarity Statistics at Each Layer}

\begin{figure}
    \centering
    \includegraphics[width=\linewidth]{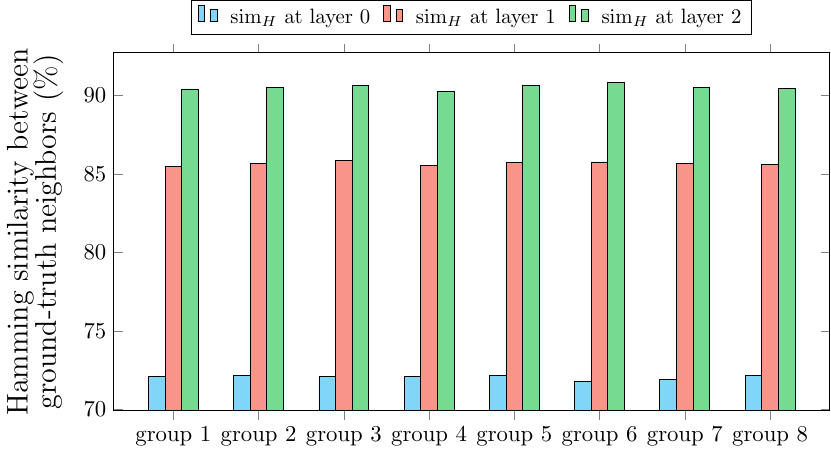}
    \caption{Layer-wise Hamming similarity statistics of ground-truth neighbors in LightGCH.}
    \label{fig:104.ground_truth_hamming_similarity}
\end{figure}

\begin{figure}
    \centering
    \includegraphics[width=\linewidth]{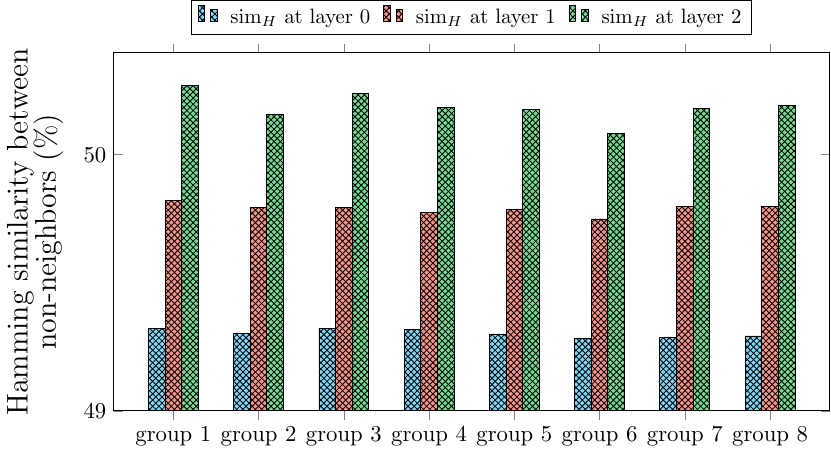}
    \caption{Layer-wise Hamming similarity statistics of non-neighbors in LightGCH.}
    \label{fig:105.neg_neighbor_hamming_similarity}
\end{figure}

Studying the generalization ability of a model based on the edge set in the test set that the model has not actually learned from is a desirable choice. On one hand, the performance on the test set is directly related to the evaluation metrics. On the other hand, since the model does not learn from the test set during training process, the statistics on it can demonstrate a model's actual generalization capability. Hence, we first obtain the hit edge set on the test set based on the fully-concatenated embeddings of LightGCH. The node pairs in the hit edge set are termed as ``ground-truth neighbors''. For node pairs that are not in the training or test sets, we term them as ``non-neighbors''. To qualitatively analyze the sign property of each layer in LightGCH, we randomly divide the source nodes in $U$ into $8$ groups and calculate the average Hamming similarity with their ground-truth neighbors. The statistical results for each layer are shown in Fig.~\ref{fig:104.ground_truth_hamming_similarity}. Using the same source node groups, we sample $2000$ non-neighbors for each source node and calculate their average Hamming similarity. The statistical results are shown in Fig.~\ref{fig:105.neg_neighbor_hamming_similarity}.

Since the average Hamming similarity of each group (Figs.~\ref{fig:104.ground_truth_hamming_similarity}-\ref{fig:105.neg_neighbor_hamming_similarity}) exhibits the same trend, we make the following qualitative analysis combining with the binarization experimental results (Fig.~\ref{fig:LightGCH_bin}):
\begin{itemize}
    \item Observing from Fig.~\ref{fig:104.ground_truth_hamming_similarity}, for actual neighbors, their shallow embeddings at layer 0 have lower Hamming similarity compared to deep embeddings because they have not undergone graph convolution operations, which lowers the overall effectiveness of embeddings to some extent. Furthermore, since the performance of shallow embeddings is not sensitive to the rescaling factor (Fig.~\ref{fig:LightGCH_bin}), the inferior performance of shallow embeddings reflects the problem of \textbf{actual neighbors having low Hamming similarity at layer 0}.
    \item Even though the Hamming similarity of deep embeddings for actual neighbors exceeds 85\% (Fig.~\ref{fig:104.ground_truth_hamming_similarity}), only using their binary embeddings does not result in significant performance gains ($\mathcal{B}_{U,V}$ in Fig.~\ref{fig:LightGCH_bin}). Moreover, after removing the rescaling factor, it causes a noticeable performance decrease of approximately 42\% (Fig.~\ref{fig:LightGCH_bin}), highlighting the excessive reliance of deep embeddings on the rescaling factor. In fact, this phenomenon precisely reflects the problem of high Hamming similarity leading to poor distinguishability of node embeddings from the perspective of sign property. Due to the high Hamming similarity at the sign level, nodes' representations may become too close, requiring additional numerical values (rescaling factor) for further discrimination. Furthermore, although graph convolution increases the Hamming similarity among neighbors, it potentially elevate the Hamming similarity of all nodes from a global perspective. The Hamming similarity of non-neighboring nodes (Fig.~\ref{fig:105.neg_neighbor_hamming_similarity}) also validates this issue: \textbf{with the increase in the number of graph convolution layers, the deep embeddings of all nodes tend to have higher Hamming similarity, resulting in a potential indistinguishability issue from the global perspective}.

    We provide a more intuitive explanation to the above analysis for deep embeddings. As shown in Fig.~\ref{fig:100.SGBGH_framework}c, we binarize the sub-embeddings of nodes in $V$, convert them into $2$-dimensional vectors using t-SNE~\cite{t-SNE}, and visualize the distribution of binary embeddings through kernel density estimation (KDE)~\cite{KDE} (with other visualization processes following the procedures in~\cite{SimGCL}). Here, the binarization operation can be understood as put embeddings into several buckets at the sign level. We preprocess embeddings with binarization to better validate that graph convolution operations elevate the overall Hamming similarity of deep embeddings in the Hamming space. Observe from Fig.~\ref{fig:100.SGBGH_framework}c that: on one hand, there are some blank spaces in the distribution of deep embeddings of LightGCH, indicating that some parts of the representation space remain unused; on the other hand, some areas in Fig.~\ref{fig:100.SGBGH_framework}c are darker in color, indicating that node embeddings may be squeezed into some corners in the Hamming space. These two phenomena can be summarized as uneven distribution of node embeddings in the deep layers of the network.
\end{itemize}

Combining the above two analyses, the issues of LightGCH can be summarized as follows: (1) low Hamming similarity of actual neighbors at layer 0; (2) high Hamming similarity of all nodes in the deep layers. To some extent, the issues with shallow and deep embeddings are even somewhat opposite, suggesting the need for different solutions at different layers.

\section{Sign-Guided Bipartite Graph Hashing Framework}

\begin{figure*}
	\centering
        \includegraphics[width=\linewidth]{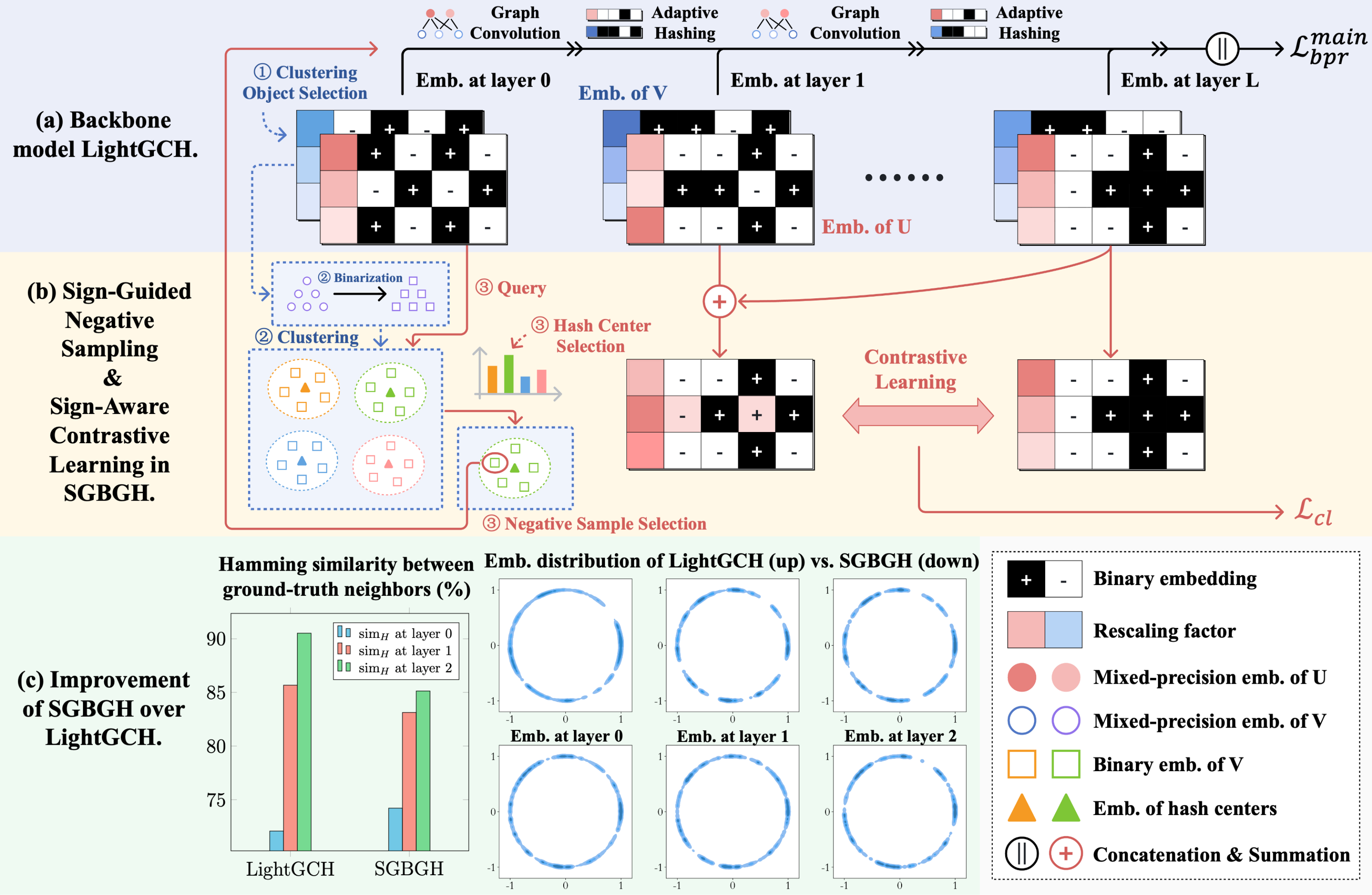}
        \caption{SGBGH framework. (a) Lightweight Convolutional Hashing backbone model LightGCH. (b) Sign-guided negative sampling and sign-aware contrastive learning in SGBGH. (c) Improvement of SGBGH over LightGCH in Hamming similarity and embedding distribution.}
	\label{fig:100.SGBGH_framework}
\end{figure*}

Here, we propose the sign-guided bipartite graph hashing framework SGBGH to address the two issues in LightGCH. Specifically, it consists of two modules: \textit{sign-guided negative sampling} and \textit{sign-aware contrastive learning}. To tackle the problem of low Hamming similarity of actual neighbors at the shallow layer, we employ a sign-guided approach for hard sample selection at layer 0. To address the issue of high Hamming similarity of all nodes at the deep layers, we utilize sign-aware contrastive learning to guide the model to learn more uniform node representations. The framework of SGBGH is illustrated in Fig.~\ref{fig:100.SGBGH_framework}.

\subsection{Sign-Guided Negative Sampling}

For the problem of low Hamming similarity of actual neighbors at layer 0, it fundamentally reflects the difficulty of mapping actual neighbors to close positions in the Hamming space. The challenge lies in that it is hard for a model to understand which nodes should be close from the perspective of the entire embedding space, when there is no additional constraints such as graph convolution. To address this issue, some works in the computer vision field suggest constructing hash centers when applying learning to hash to image classification tasks. 
Their main idea is to build a hash center for each classification category, so that the hash embedding of each image should be close to the center of its category while being far from the centers of other categories~\cite{CgAT,PSLDH}. The modeling approach of using hash centers as an intermediate is highly instructive to learning to hash: if we incorporate image classification tasks into the framework of bipartite graph embedding, images and classification categories can be mapped to source nodes in $U$ and destination nodes in $V$ in the bipartite graph, respectively. Although there is no explicit classification categories in the bipartite graph, we can still construct pseudo hash centers through clustering algorithms to guide the node embeddings to be closer or further apart.

Specifically, we use these pseudo hash centers to guide the sampling method $N(\cdot)$ in Eq.~\eqref{eq:hash_loss_main}. First, clustering is performed on the destination nodes in $V$ to obtain $|C|$ pseudo hash centers and the assignment of hash centers for each node $v \in V$. Then, according to the query node $u \in U$, sampling is conducted within the hash centers, followed by sampling within the corresponding node set of the hash centers to obtain the negative sample $v'$ for training.

The rationality of the above approach is analyzed as follows. Intuitively, the embeddings in the Hamming space have limited permutations. If a node's embedding is very close to other embeddings at the sign level, the distinguishability brought by the sign will be reduced, leading to the problem of excessive influence of the rescaling factor on performance (Fig.~\ref{fig:LightGCH_bin}). Only by maintaining a distance from other embeddings as much as possible can the embeddings distribute everywhere as much as possible in the Hamming space. For a node $u \in U$, selecting negative samples with too low Hamming similarity will make them easily distinguishable at the sign level, which does not provide much help for training. In essence, the negative sampling problem lies in selecting samples that can maximize the training benefits. A feasible approach is to select hard negative samples with relatively high Hamming similarity, as this can provide more useful training information in the Hamming space.

Based on the aforementioned idea, we propose a sign-guided hard negative sampling method, which consists of three key steps: \textit{clustering object selection}, \textit{hash center calculation}, and \textit{sampling}.

\subsubsection{Clustering Object Selection}

Although our idea is to cluster the nodes in $V$, each node $v \in V$ is concatenated by sub-embeddings at each layer, i.e., $\mathbf{q}_{v} = \mathbf{q}_{v}^{(0)} \Big{\|} \mathbf{q}_{v}^{(1)} \Big{\|} \cdots \Big{\|} \mathbf{q}_{v}^{(L)}$, choosing which part of the embedding for clustering is a concern. We choose the shallow embedding $\mathbf{q}_{v}^{(0)}$ at layer 0 based on four considerations: (1) In terms of the quality of sub-embeddings across layers, the sub-embedding at layer 0 performs the worst (Fig.~\ref{fig:LightGCH_bin}), conducting negative sampling for it is the most direct solution. (2) Since graph convolution operations tend to increase the Hamming similarity of deep embeddings (Figs.~\ref{fig:104.ground_truth_hamming_similarity}-\ref{fig:105.neg_neighbor_hamming_similarity}), hard negatives sampled from layer 0 may also be more similar at the deep layers with the assistance of graph convolution. (3) The statistical results of Hamming similarity across layers in LightGCH (Figs.~\ref{fig:104.ground_truth_hamming_similarity}-\ref{fig:105.neg_neighbor_hamming_similarity}) indicate that the distributions of embeddings across layers are different, so treating each part/layer of the concatenated sub-embeddings equally for clustering may not be suitable. (4) If fully-concatenated embeddings are used for clustering, the dimensionality is higher, leading to increasing computational costs during clustering. The approach for selecting clustering objects is shown in Fig.~\ref{fig:100.SGBGH_framework}b-\ding{172}.

\subsubsection{Hash Center Calculation}

Before selecting negative samples based on Hamming similarity, it is necessary to partition candidate samples into clusters using clustering algorithms such as K-means, with the center of each cluster serving as a hash center. Before clustering, we first binarize the shallow embedding of each destination node $v \in V$, obtaining $\mathbf{b}_{v}^{(0)} = \operatorname{sign}(\mathbf{q}_{v}^{(0)})$. We use  binarization based on the observation that removing the rescaling factor from node embeddings in $V$ results in a significant performance decrease in most cases (Fig.~\ref{fig:LightGCH_bin}). This indicates that the binary embedding of $V$ ($\mathcal{B}_V$) is more likely to determine the lower limit of embedding performance. If the rescaling factor is removed before clustering, it can eliminate the influence of numerical value information on clustering. Consequently, the magnitude of each dimension of the clustering center (hash center) reflects the corresponding quantity of signs in the cluster members. Then, the belongingness of node $v$ to a cluster mainly depends on its sign matching with the clustering center. Subsequently, we employ K-means clustering based on the binary embeddings at layer 0 to obtain the set of hash centers $C=\{c_1, c_2, \cdots, c_{|C|}\}$. The two steps for hash center computation are shown in Fig.~\ref{fig:100.SGBGH_framework}b-\ding{173}.

\subsubsection{Sampling}

Given a source node $u \in U$, the purpose of hard negative sampling is to find negative samples with relatively high Hamming similarity. Since we have already obtained the hash centers, we can approximate the relevance between $u$ and each hash center based on their dot product. In brief, we select a hash center based on the dot product values, and then sample from the nodes assigned to that center to obtain a negative sample. Therefore, the sampling process can be divided into two steps: \textit{hash center selection}, and \textit{negative sample selection}. Their implementations are as follows: 

\begin{enumerate}[i)]
    \item Hash Center Selection: Given $u \in U$, the probability of selecting each hash center $c_i$ is calculated as $Pr(c_i|u) = \frac{\operatorname{exp}({\mathbf{q}_u^{(0)}}^\mathnormal{T} \bf{c}_\mathnormal{i})}{\sum_{c_j \in C } \operatorname{exp}({\mathbf{q}_u^{(0)}}^\mathnormal{T} \bf{c}_\mathnormal{j})}$.
    \item Negative Sample Selection: After selecting the hash center, we uniformly sample negative samples from the nodes assigned to that hash center.
\end{enumerate}

The sampling process is shown in Fig.~\ref{fig:100.SGBGH_framework}b-\ding{174}.

\subsection{Sign-Aware Contrastive Learning}

The problem of high Hamming similarity in deep embeddings can be attributed to the embeddings being closer in terms of signs, leading to indistinguishability. Only when the embeddings distribute more uniformly or evenly, the limited Hamming space can be better utilized for representation. To improve the uniformity of embeddings, related research has discussed that contrastive learning  happens to provide such a characteristic~\cite{SimGCL,XSimGCL}.

Although there have been many works using contrastive learning to improve the quality of bipartite graph embedding, (1) they are tailored towards floating-point embeddings, which differ from learning to hash; (2) the final embedding representation of LightGCH is obtained by concatenation rather than average, adapting to learning to hash needs further consideration; (3) many related works use augmentation methods like edge dropout and node dropout for contrastive learning, but some augmentation methods have been found to have limited help in improving performance. Therefore, it is worth exploring a simpler contrastive method to improve the uniformity of node embeddings, and thereby enhance the effectiveness of BGH.

In fact, from the statistics of Hamming similarity in Figs.~\ref{fig:104.ground_truth_hamming_similarity}-\ref{fig:105.neg_neighbor_hamming_similarity}, we observe that the Hamming similarity at the last layer of LightGCH is the highest. Therefore, an intuitive idea to improve embedding uniformity is to perform contrastive learning for the last layer. For source nodes, the contrastive loss is as follows:
\begin{equation}  \label{eq:hash_cl_last}
\mathcal{L}_{cl}^{u}=\sum_{u_i \in U}-\log \frac{\exp \left(\operatorname{sim} \left(\mathbf{q}_{u_i}^{(L)}, \mathbf{q}_{u_i}^{(L)}\right) / \tau\right)}{\sum_{u_j \in \mathfrak{B}} \exp \left(\operatorname{sim}\left(\mathbf{q}_{u_i}^{(L)}, \mathbf{q}_{u_j}^{(L)}\right) / \tau\right)} ,
\end{equation}
where $\mathfrak{B}$ represents the set of nodes of the same type in a training batch, $\mathbf{q}_{u_i}^{(L)}$ and $\mathbf{q}_{u_j}^{(L)}$ are the mixed-precision embeddings of nodes $u_i$ and $u_j$ at the last layer respectively, $\tau$ is the temperature for contrastive learning, and $\operatorname{sim}(\cdot, \cdot)$ is a similarity measure function, which is defined as the cosine similarity in SGBGH.

By optimizing Eq.~\eqref{eq:hash_cl_last}, SGBGH is able to conduct sign-aware contrastive learning:
\begin{equation}
\begin{aligned}
\mathcal{L}_{cl}^{u} &= \sum_{u_i \in U}-\log \frac{\exp \left(\operatorname{sim} \left(\mathbf{q}_{u_i}^{(L)}, \mathbf{q}_{u_i}^{(L)}\right) / \tau\right)}{\sum_{u_j \in \mathfrak{B}} \exp \left(\operatorname{sim}\left(\mathbf{q}_{u_i}^{(L)}, \mathbf{q}_{u_j}^{(L)}\right) / \tau\right)}  \\
    &= \sum_{u_i \in U}-\log \frac{\exp \left( 1 / \tau\right)}{\sum_{u_j \in \mathfrak{B}} \exp \left(\operatorname{sim}\left(\alpha_{u_i}^{(L)}\mathbf{b}_{u_i}^{(L)}, \alpha_{u_j}^{(L)}\mathbf{b}_{u_j}^{(L)}\right) / \tau\right)}  \\
    &= \sum_{u_i \in U}-\log \frac{\exp \left( 1 / \tau\right)}{\sum_{u_j \in \mathfrak{B}} \exp \left( \frac{\alpha_{u_i}^{(L)} \alpha_{u_j}^{(L)} {\mathbf{b}_{u_i}^{(L)}}^T \mathbf{b}_{u_j}^{(L)}}{\sqrt{d \cdot \left({\alpha_{u_i}^{(L)}}\right)^2 } \cdot \sqrt{d \cdot \left({\alpha_{u_j}^{(L)}}\right)^2 } \cdot \tau} \right)}  \\
    &= \sum_{u_i \in U}-\log \frac{\exp \left( 1 / \tau\right)}{\sum_{u_j \in \mathfrak{B}} \exp \left( \frac{\alpha_{u_i}^{(L)} \alpha_{u_j}^{(L)} {\mathbf{b}_{u_i}^{(L)}}^T \mathbf{b}_{u_j}^{(L)}}{d \cdot \alpha_{u_i}^{(L)} \cdot  \alpha_{u_j}^{(L)} \cdot \tau} \right)}  \\
    &= \sum_{u_i \in U}-\log \frac{\exp \left( 1 / \tau\right)}{\sum_{u_j \in \mathfrak{B}} \exp \left( {\mathbf{b}_{u_i}^{(L)}}^T \mathbf{b}_{u_j}^{(L)} / \left( d \tau \right) \right)}.  \\
\end{aligned}
\end{equation}
Therefore, constraining the cosine similarity of the mixed-precision embeddings is equivalent to constraining the dot product of their binary embeddings.

The above approach performs well when $L=1$. However, when stacking more layers ($L \geq 2$), the performance improvement is not significant. We conjecture that the graph convolution itself tends to make the deep embeddings of nodes more similar. However, applying contrastive learning only to the last layer may disrupt this property, making the intermediate layers dissimilar from the last layer. Therefore, we modify Eq.~\eqref{eq:hash_cl_last} by summing the deep embeddings and contrasting the summation result with the embeddings at the last layer:
\begin{equation}  \label{eq:hash_cl_exclude0}
\mathcal{L}_{cl}^{u}=\sum_{u_i \in U}-\log \frac{\exp \left(\operatorname{sim} \left(\mathbf{q}_{u_i}^{(L)}, \mathbf{e}_{u_i}^{*}\right) / \tau\right)}{\sum_{u_j \in \mathfrak{B}} \exp \left(\operatorname{sim}\left(\mathbf{q}_{u_i}^{(L)}, \mathbf{e}_{u_j}^{*}\right) / \tau\right)} ,
\end{equation}
where $\mathbf{e}_{u_i}^{*} = \sum\nolimits_\mathnormal{l\rm{=1}}^\mathnormal{L}  w_{l} \mathbf{q}_{u_i}^{(l)}$ is the result of the summation of sub-embeddings at the deep layers, and $w_{l}$ can be a hyperparameter or a learnable variable, which is set as a constant in this paper for each layer following~\cite{LightGCN,LayerGCN}. 
Compared to Eq.~\eqref{eq:hash_cl_last} where the similarity is only calculated for the embeddings at the last layer and its value in the numerator is always $1$, Eq.~\eqref{eq:hash_cl_exclude0} introduces some additional constraints: the similarity in the numerator depends on the summation result of the last layer and the preceding layers, thereby constraining the summation result to be as close as possible to the embedding at the last layer.

Last, since a bipartite graph contains both source nodes ($U$) and destination nodes ($V$), most existing works perform contrastive learning for both types of nodes simultaneously~\cite{SGL,SimGCL}. However, in our experiments, we find that while conducting contrastive learning to either side (for $U$ or $V$ exclusively) can improve performance, the greater performance gain usually comes from contrastive learning applied to the source nodes. Therefore, the final contrastive loss is defined as:
\begin{equation}  \label{eq:hash_cl_u}
\mathcal{L}_{cl} = \mathcal{L}_{cl}^{u}  .
\end{equation}

The process of contrastive learning is shown in Fig.~\ref{fig:100.SGBGH_framework}b.

\subsection{Multi-Objective Optimization}

The final loss function of SGBGH is defined as:
\begin{equation}  \label{eq:hash_loss}
    \mathcal{L} = \mathcal{L}_{bpr}^{main} + \gamma \mathcal{L}_{cl} + \beta_0 \mathcal{L}_{bpr}^0 + \beta_1 \mathcal{L}_{bpr}^{conv} + \lambda \| \Theta \|^2  ,
\end{equation}
where $\mathcal{L}_{bpr}^{main}$ is the BPR loss for the fully-concatenated embeddings in Eq.~\eqref{eq:hash_loss_main}, $\mathcal{L}_{cl}$ is the contrastive loss. Besides, since SGBCH performs negative sampling based on the embeddings at layer 0, an additional BPR loss $\mathcal{L}_{bpr}^0$ is introduced to constrain the shallow embeddings. Similarly, since SGBCH performs contrastive learning for deep embeddings, an additional BPR loss $\mathcal{L}_{bpr}^{conv}$ is used to constrain the concatenated sub-embeddings from layers $1, 2, \cdots, L$. $\gamma$, $\beta_0$, $\beta_1$ and $\lambda$ are all hyperparameters.

\section{Complexity Analysis}

\subsection{Training Time Complexity}

The time complexity of SGBGH is mainly dependent on graph convolutional hashing, contrastive learning and three BPR loss functions. In each training batch, graph convolutional hashing needs $O(|E|Ld)$ time, contrastive learning needs $O(BMd)$ time, where $B$ and $M$ denote batch size and the number of nodes in contrast respectively, and three BPR loss functions need $O(B(L+1)d)$ time.

\subsection{Storage Overhead}

When the dimensionality is $d$ per layer and each floating-point number occupies $32$ bits, the $L$-layer SGBGH will require $(|U|+|V|)(L+1)(d+32)$ bits of storage. In comparison to floating-point embeddings (which require 
$32(|U|+|V|)d$ bits of storage), their ratio of storage occupation is 
$\frac{(L+1)(d+32)}{32d}$. Generally, $L \leq 3$ and $d \geq 64$, therefore, embeddings obtained through learning to hash will achieve a $5\times$ compression rate at least.

\subsection{Inference Overhead}

We use FLOPs (floating-point operations) and BOPs (binary operations) to measure the inference overhead. Compared to the floating-point embeddings which need $O(|U| \cdot |V| \cdot d)$ FLOPs, the $L$-layer SGBGH needs $O(|U| \cdot |V| \cdot (L+1))$ FLOPs and $O(|U| \cdot |V| \cdot (L+1)d)$ BOPs, where most floating-point operations are converted to binary operations, while introducing only a small amount of additional floating-point operations.

\section{Experimental Study}

We conduct model training for all methods on a Linux server with an Intel(R) Xeon(R) CPU E5-2680 v4 @ 2.40GHz CPU, 62GB RAM and a GeForce RTX 3090 GPU (24GB). The experiments for online inference are conducted on a MacBook Pro equipped with an Apple M1 Pro chip and 32GB RAM.

\subsection{Experiment Setup}

\subsubsection{Datasets} We use five commonly used datasets in bipartite graph hashing: MovieLens-1M (referred to as Movie), Gowalla, Pinterest, Yelp and Amazon-Book. The statistics of these datasets are reported in Table~\ref{tab:SGBGH_data}. All these datasets are collected from~\cite{BGCH}.

\begin{table}
    \centering
    \caption{The statistics of datasets.}
    \label{tab:SGBGH_data}
    \begin{tabular}{c|c|c|c|c}
        \toprule
        \textbf{Datasets} & {$|U|$} & {$|V|$} & {$|E|$} & {\textbf{Density}} \\
        \midrule
        Movie & 6,040& 3,952 & 1,000,209 & 0.04190 \\
	Gowalla & 29,858& 40,981 & 1,027,370 & 0.00084 \\
	Pinterest & 55,186& 9,916 & 1,463,556 & 0.00267 \\
	Yelp & 31,668& 38,048 & 1,561,406 & 0.00130 \\
	Amazon-Book & 52,643& 91,599 & 2,984,108 & 0.00062 \\
        \bottomrule
    \end{tabular}
\end{table}

\subsubsection{Implementation Details} 

We implement SGBGH based on the PyTorch framework. Model parameters are initialized using the Xavier method~\cite{Xavier}, and the learning rate is tuned within $\{0.0005, 0.001\}$. Adam optimizer~\cite{Adam} is employed for model optimization during training. The batch size is set to $B = 4096$. The number of hash centers $|C|$ is set to $64$. For contrastive learning, the temperature $\tau$ is set to $0.2$, and the weight coefficient is searched in $\gamma \in [0,1]$. For the two additional BPR loss functions, their weights are tuned by $\beta_0,\beta_1 \in \{0, 1, 2\}$. The L2 regularization coefficient $\lambda$ is set to $0.001$. Additionally, to ensure fair comparison, the embedding dimensionality $d$ is set to $64$ for all models, and the number of graph convolution layers $L$ is set to $2$.

\subsubsection{Baselines}

To comprehensively compare with existing bipartite graph hashing methods, we select recent five baseline models based on deep learning for comparison. They can be categorized into two groups:

\begin{itemize}
    \item End-to-end hashing methods: LightGCH (a lightweight model proposed by us), BGCH~\cite{BGCH}, HS-GCN~\cite{HS-GCN} and HashGNN~\cite{HashGNN}. For HashGNN, we follow~\cite{BGCH} to use $\rm{HashGNN_h}$ to denote the vanilla version with hard binary encoding proposed in~\cite{HashGNN}, and use $\rm{HashGNN_s}$ to denote its approximated version.
    \item Two-phase quantization-based method: BiGeaR~\cite{BiGeaR}.
\end{itemize}

\subsection{Top-$K$ Hamming Space Search}

\begin{figure}
	\centering
        \includegraphics[width=\linewidth]{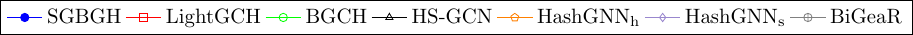}\\
	\begin{minipage}[t]{\linewidth}
			\centering
	\subfloat[Gowalla]{
			\centering
                \includegraphics[width=0.31\linewidth]{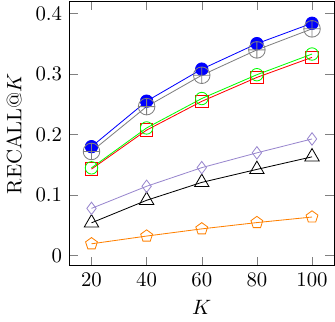}
		\label{fig:hash_topk_gowalla}
	}
        \subfloat[Yelp]{
			\centering
                \includegraphics[width=0.31\linewidth]{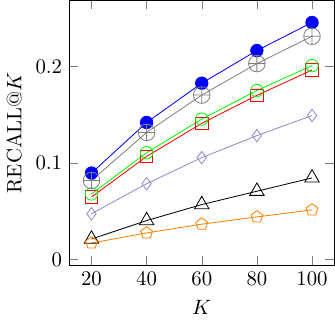}
		\label{fig:hash_topk_yelp}
	}
         \subfloat[Amazon-Book]{
			\centering
                \includegraphics[width=0.31\linewidth]{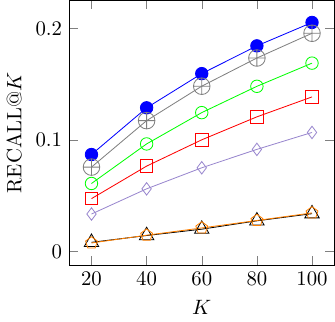}
		\label{fig:hash_topk_book}
	}
	\centering
	\caption{Top-$K$ search quality with $K$ in \{20, 40, 60, 80, 100\}.} \label{fig:hash_topk}
\end{minipage}%
\end{figure}

To evaluate the quality of the learned hash embeddings, we choose Top-$K$ search as the downstream task of BGH to assess the effectiveness of Hamming space retrieval. All datasets are divided into training and test sets in a ratio of $8:2$. To evaluate the model's fine-to-coarse ranking capability, we set $K$ to $100$. In Table~\ref{tab:hash_search}, we report the $\rm{RECALL@20_{100}}$ and $\rm{NDCG@20_{100}}$ metrics for all methods. Furthermore, in Fig.~\ref{fig:hash_topk}, we plot the overall RECALL curves of $\{20, 40, 60, 80, 100\}$ of Top-100 on Gowalla, Yelp and Amazon-Book datasets for all methods.

\begin{table*}
        \centering
	\caption{Results of Recall@20 and NDCG@20 in Top-100 retrieval. The best result is highlighted in bold, the second best is underlined, and the third best is  doubly underlined.}
	\label{tab:hash_search}
	\begin{tabular}{c|cc|cc|cc|cc|cc}
        \toprule
	\multirow{2}{*}{\textbf{Method}} & \multicolumn{2}{c|}{\textbf{Movie}} & \multicolumn{2}{c|}{\textbf{Gowalla}} & \multicolumn{2}{c|}{\textbf{Pinterest}} & \multicolumn{2}{c|}{\textbf{Yelp}} & \multicolumn{2}{c}{\textbf{Amazon-Book}}\\
		{} & RECALL & NDCG & RECALL & NDCG & RECALL & NDCG & RECALL & NDCG & RECALL & NDCG \\
		\midrule
		SGBGH & \underline{0.212} & \underline{0.327} & \textbf{0.180} & \textbf{0.136} & \textbf{0.148} & \textbf{0.099} & \textbf{0.090} & \textbf{0.075} & \textbf{0.087} & \textbf{0.077}\\
            LightGCH & 0.114 & 0.191 & 0.143 & 0.108 & \underline{\underline{0.124}} & \underline{\underline{0.082}} & 0.065 & 0.053 & 0.048 & 0.041\\
		BGCH & \underline{\underline{0.157}} & \underline{\underline{0.257}} & \underline{\underline{0.145}} & \underline{\underline{0.109}} & 0.122 & 0.081 & \underline{\underline{0.068}} & \underline{\underline{0.056}} & \underline{\underline{0.061}} & \underline{\underline{0.051}} \\
		HS-GCN & 0.135 & 0.219 & 0.054 & 0.036 & 0.066 & 0.041 & 0.021 & 0.016 & 0.008 & 0.007 \\
		$\rm{HashGNN_h}$ & 0.098 & 0.162 & 0.019 & 0.012 & 0.040 & 0.024 & 0.017 & 0.015 & 0.008 & 0.006 \\
		$\rm{HashGNN_s}$ & 0.152 & 0.240 & 0.078 & 0.057 & 0.101 & 0.067 & 0.048 & 0.038 & 0.034 & 0.028 \\
		\midrule
  		BiGeaR & \textbf{0.227} & \textbf{0.345} & \underline{0.172} & \underline{0.128} & \underline{0.145} & \underline{0.098} & \underline{0.082} & \underline{0.068} & \underline{0.076} & \underline{0.065} \\
		\bottomrule
	\end{tabular}
\end{table*}

In terms of the overall embedding quality, SGBGH achieves the best results on all datasets except for Movie. Compared to the backbone model LightGCH, SGBGH improves the RECALL metric by 19\% - 86\% and improves the NDCG metric by 21\% - 88\%. These significant improvements validate the effectiveness of the advanced sign-guided bipartite graph hashing framework. Furthermore, we draw the following conclusions: (1) As a backbone model for SGBGH, LightGCH achieves comparable results to the state-of-the-art end-to-end hashing method BGCH~\cite{BGCH} on three datasets, indicating that the graph convolutional hashing modeling approach helps generate high-quality hash embeddings. (2) Compared to LightGCH, SGBGH mainly incorporates two sign-guided methods for improvement, optimizing the shallow and deep layers of the network through negative sampling and contrastive learning, respectively, leading to a significant improvement in model performance. This demonstrates the effectiveness of sign guidance for bipartite graph hashing. (3) More specifically, the improvements of SGBGH over LightGCH are reflected at each layer of the network. We visualize the actual neighbors' Hamming similarity and embedding distribution for nodes in $V$ at different layers for demonstration, as shown in Fig.~\ref{fig:100.SGBGH_framework}c. From the perspective of Hamming similarity, the similarity of embeddings at layer 0 is higher compared to LightGCH, while the similarity at layers 1 and 2 is lower, alleviating the issues of the backbone model. From the perspective of embedding distribution, SGBGH learns a more uniform embedding distribution compared to LightGCH, with fewer blank spaces and higher utilization of the embedding space. Therefore, even after binarization at all layers, it still maintains high embedding quality (Fig.~\ref{fig:BGCH_bin}).

From the perspective of the embedding form, SGBGH, BGCH~\cite{BGCH}, BiGeaR~\cite{BiGeaR} and LightGCH all learn mixed-precision embeddings, which significantly outperform methods that only learn binary embeddings (HS-GCN~\cite{HS-GCN} and $\rm{HashGNN_h}$~\cite{HashGNN}) and methods that learn continuous embeddings ($\rm{HashGNN_s}$~\cite{HashGNN}) in terms of hashing performance. This is because mixed-precision embeddings not only contain binary embeddings but also retain a small amount of floating-point scaling factors for each node. Compared to binary embeddings with limited permutations in the Hamming space, mixed-precision embeddings are more expressive. Additionally, although $\rm{HashGNN_s}$ partially retain the expressiveness of floating-point embeddings by replacing binary values with original continuous values using Bernoulli random variables, it uses earlier graph convolutional networks~\cite{GraphSAGE} as its backbone model, which may not be as effective as the latest lightweight graph convolutional networks~\cite{BGCH,BiGeaR,LightGCN}.

From the perspective of the training method, the quantization-based method BiGeaR~\cite{BiGeaR} surpasses all end-to-end methods except SGBGH. This is because BiGeaR's teacher model is a floating-point model, and the superiority of its student model can be attributed to several factors: (1) The floating-point layer-wise graph convolutional propagation does not incur precision loss to the teacher model. (2) The student model's weights are initialized using the teacher's weights. (3) The student model distills the ranking knowledge from the teacher model layer by layer during training. All these three factors indicate that much of BiGeaR's performance gain comes from stronger external knowledge. In contrast, it is hard for end-to-end methods to improve performance through techniques such as reducing precision loss, extracting weight data, and distilling knowledge across layers. This results in their generally inferior performance. However, SGBGH still outperforms BiGeaR on four datasets. Additionally, compared to the state-of-the-art end-to-end method BGCH, SGBGH achieves at least 21\% improvement in both RECALL and NDCG on all datasets. This clearly demonstrates the merits of the proposed sign-guided approach for bipartite graph hashing, which emphasizes the importance of the sign property in learning to hash and makes improvements at both shallow and deep layers of the network to mitigate the corresponding shortcomings.

SGBGH exhibits more pronounced advantages on sparse graphs. On Amazon-Book dataset, its RECALL is 43\% higher than BGCH and 14\% higher than BiGeaR. On Yelp dataset, its RECALL surpasses BGCH by 32\% and BiGeaR by 10\%. Since sparse graphs often have a larger number of nodes, utilizing SGBGH for approximate Top-$K$ search can yield greater benefits compared to BGCH and BiGeaR under equivalent storage overhead.

Additionally, as shown in Fig.~\ref{fig:hash_topk}, as $K$ varies from 20 to 100, SGBGH consistently outperforms the state-of-the-art end-to-end hashing method BGCH and the quantization-based method BiGeaR on three datasets in terms of RECALL and NDCG. On Amazon-Book dataset, SGBGH's results for RECALL@\{20, 40, 60, 80, 100\} are respectively 43\%, 33\%, 28\%, 25\%, and 22\% higher than BGCH, and 14\%, 9\%, 8\%, 6\%, and 5\% higher than BiGeaR. These results demonstrate the superior and stable hashing performance of SGBGH. Besides, as $K$ increases, the improvement rate decreases, with the highest improvement rate observed under RECALL@20. This indicates that compared to existing methods, SGBGH can provide finer-grained retrieval performance in the Hamming space. Furthermore, SGBGH offers the advantage of shorter training time compared to BGCH and BiGeaR, which will be further analyzed in Section~\ref{sec:hash_resource}.

\subsection{Resource Consumption} \label{sec:hash_resource}

We select baseline models BGCH, $\rm{HashGNN_s}$, and BiGeaR with superior hashing performance for resource consumption comparison, including overall training time, embedding storage overhead, and online inference overhead.

\subsubsection{Training Time}

\begin{figure}
    \centering
    \includegraphics[width=\linewidth]{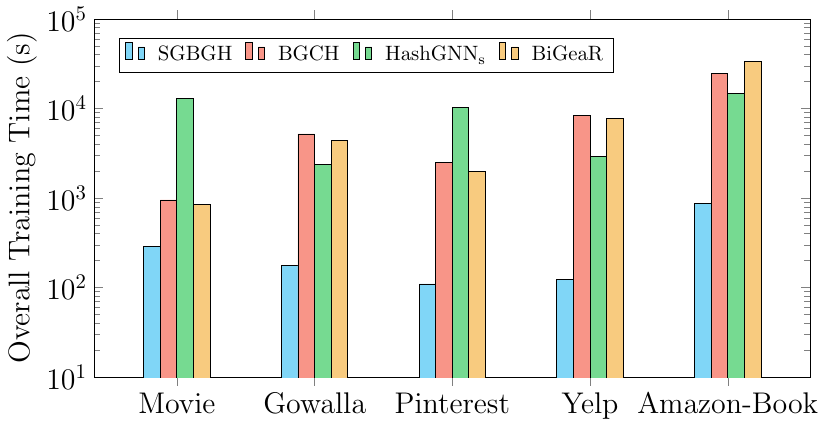}
    \caption{Overall training time (s).}
    \label{fig:110.hash_overall_training_time}
\end{figure}

The overall training time of competitor methods is shown in Fig.~\ref{fig:110.hash_overall_training_time}. Observe that SGBGH has the shortest training time across all five datasets. Compared to $\rm{HashGNN_s}$, SGBGH achieves up to $93\times$ speedup. This is because $\rm{HashGNN_s}$ adopts earlier graph convolutional networks~\cite{GraphSAGE}, which often require extensive neighborhood sampling for nodes' context construction, resulting in relatively higher time consumption. In contrast, the latest graph convolutional networks typically only require pre-computating and storing the normalized adjacency matrix without additional overhead for context construction. Additionally, compared to the state-of-the-art methods BGCH and BiGeaR, SGBGH achieves up to $68\times$ and $62\times$ training speedups, respectively. In summary, as an end-to-end bipartite graph hashing method, SGBGH demonstrates faster training speed and superior performance on most datasets, thus exhibiting significant advantages in training efficiency.

\subsubsection{Storage Overhead and Inference Overhead}

\begin{figure}
	\centering
	\begin{minipage}[t]{\linewidth}
			\centering
	\subfloat[Storage overhead]{
			\centering
                \includegraphics[width=0.4\linewidth]{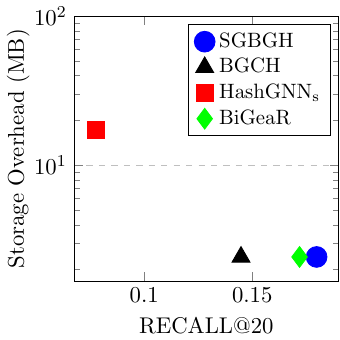}
		\label{fig:hash_perf_vs_storage}
	}
        \subfloat[Inference overhead]{
			\centering
                \includegraphics[width=0.4\linewidth]{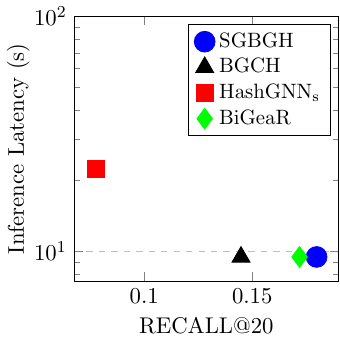}
		\label{fig:hash_perf_vs_inference}
	}
	\centering
	\caption{Storage and inference overheads on Gowalla dataset.} \label{fig:hash_perf_vs_storage_inference}
\end{minipage}%
\end{figure}

In terms of storage overhead and online inference time, since SGBGH shares the same embedding form as BGCH and BiGeaR, their storage and inference overheads are also the same. Besides, since the advantages of mixed-precision embeddings over $\rm{HashGNN_s}$ in storage and inference have already been demonstrated in experiments of~\cite{BGCH,BiGeaR}, we focus on how different methods balance embedding quality with storage/inference overhead in this paper. We conduct experiments on Gowalla dataset and compare SGBGH with BGCH, $\rm{HashGNN_s}$ and BiGeaR. For online inference scenario, we follow the approach outlined in~\cite{BiGeaR} and randomly construct $1000$ queries to measure the inference time. Figs.~\ref{fig:hash_perf_vs_storage}-\ref{fig:hash_perf_vs_inference} show the trade-offs between embedding performance and storage overhead, as well as embedding performance and inference overhead for each method.

(1) In terms of storage overhead, methods that learn mixed-precision embeddings not only save 86\% of storage compared to $\rm{HashGNN_s}$ on Gowalla dataset but also bring an average performance improvement of 113\%. This validates the effectiveness of mixed-precision embeddings for bipartite graph hashing. Among these methods, SGBGH performs the best. With the same storage overhead, it exhibits stronger embedding expressiveness compared to BGCH and BiGeaR.

(2) From the perspective of inference overhead, the dot product computation of mixed-precision embeddings can be accelerated using bitwise operations. Therefore, SGBGH, BGCH and BiGeaR have faster inference speed compared to $\rm{HashGNN_s}$ due to their mixed-precision embedding form. This is because $\rm{HashGNN_s}$ randomly replaces binary values in embeddings with floating-point values, making it difficult to accelerate computation using bitwise operations. Overall, the mixed-precision embedding form is effective for retrieval-based online inference scenarios.

\section{Conclusion}

In this paper, we propose a novel sign-guided framework SGBGH for bipartite graph hashing. 
Specifically, we first construct a lightweight graph convolutional hashing model LightGCH based on the state-of-the-art end-to-end method BGCH~\cite{BGCH}. 
By analyzing the Hamming similarity of ground-truth neighbors and non-neighbors in the bipartite graph, combining with the performance trends at each layer under different binarization scenarios, we identify two main issues in LightGCH: low Hamming similarity of actual neighbors at the shallow layer and high Hamming similarity of all nodes at the deep layers. Then we propose SGBGH for improvement. For the shallow layer issue, we propose sign-guided negative sampling to improve the Hamming similarity of neighboring nodes. For the deep layer issue, we propose sign-aware contrastive learning to guide nodes to learn a more uniform distribution. Experimental results validate the advanced embedding quality of SGBGH, outperforming BGCH by at least 21\% and outperforming LightGCH by at least 19\% across all datasets.

\bibliographystyle{IEEEtran}
\bibliography{ref}

\end{document}